\def\doverd#1#2{\frac{\partial #1}{\partial #2}}
\def\doverdt#1{\frac{\partial #1}{\partial t}}
\def\doverdr#1{\frac{\partial #1}{\partial r}}
\def\subscr#1{_{\rm #1}}

\def\csound{c\subscr{s}}
\def\Jdot{\dot{J}}
\def\Mdot{\dot{M}}
\def\Msol{\rm M_{\odot}}
\def\Mstar{M\subscr{*}}
\def\Omk{\Omega\subscr{k}}

\def\Tstar{T\subscr{*}}
\def\Trp{T_{r \varphi}}
\def\Trpo{T^0_{r \varphi}}

\documentstyle[psfig]{lamuphys}

\def\amin{\ifmmode ^{\prime}\else$^{\prime}$\fi}
\def\asec{\ifmmode ^{\prime\prime}\else$^{\prime\prime}$\fi}

\begin{document}        

\title{Causal Viscosity in Accretion Disc Boundary Layers}

\titlerunning{Causal Viscosity}

\author{W. Kley\inst{1}, J.~C.~B. Papaloizou\inst{2}}

\institute{Max-Planck-Society, Research Unit Gravitational Theory,
Universit\"at Jena, Max-Wien-Platz 1, D-07743 Jena, Germany
\and
Astronomy Unit, School of Mathematical Sciences,
Queen Mary \& Westfield College, Mile End Road, London E1 4NS, UK}

\maketitle

\index{KLey, W.|uu}

\glossary{accretion disk}
\glossary{boundary layer}
\glossary{viscosity, causal}

\noindent{\bf Abstract:}
The structure of the boundary layer region
between the disc and a comparatively slowly rotating star is studied
using a causal prescription for viscosity.
The vertically integrated viscous stress
relaxes towards its equilibrium value on a relaxation timescale $\tau$,
which naturally yields a finite speed of propagation for viscous information.
For a standard $\alpha$ prescription
with $\alpha$ in the range $0.1-0.01,$ and ratio of viscous speed to sound speed
 in the range $0.02-0.5,$ details in the boundary layer are
strongly affected by the causality constraint.  We  study both steady state
polytropic models and time dependent models, taking into account energy
dissipation and transport.
Steady state solutions are always subviscous with a variety of $\Omega$ profiles
 which
may exhibit near discontinuities.
 For $\alpha =0.01$ and small viscous speeds, the
boundary layer adjusted to a near steady state. A long wavelength
oscillation generated by viscous overstability could be seen
\glossary{viscous overstability}
at times near the outer
boundary. Being confined there, the boundary layer remained
almost stationary. However, for $\alpha =0.1$ and large viscous speeds,
short wavelength disturbances were seen throughout which could significantly
affect  the
power output in the boundary layer. This could be potentially
 important
in producing time dependent behaviour in accreting systems such as CVs and
protostars.

\section{Introduction}
The boundary layer region between a star and accretion disc is of fundamental
importance for non-magnetic accreting systems. This is because up to half
the total accretion energy may be liberated over a relatively small
scale in this region.
(Lynden-Bell and Pringle 1974, Pringle 1977). Consequently, the angular
velocity changes rapidly from a near  Keplerian value to a smaller
value associated with the accreting star on a scale length that is expected
to be comparable to the pressure scale height of the slowly rotating star.

In a thin Keplerian disc, the inflow velocity is generally highly subsonic.
 However, in the boundary layer where
the gradients increase the radial infall
velocity may become large, reaching supersonic values,
if an unmodified viscosity prescription appropriate to the outer disc is used
(see Papaloizou and Stanley 1986; Kley 1989, Popham and Narayan 1992).
In this case, it has been argued (Pringle 1977) that the star would
lose causal connection with the outer parts of the disc so that information
 about the inner boundary conditions could not  be communicated outward.
In order to prevent such a situation, the viscosity prescription should be
modified so as to prevent unphysical communication of information.
 Various
approaches that limit the viscosity in the vicinity of the star (thus reducing
the radial inflow velocity) have been suggested (Papaloizou \& Stanley 1986,
Popham \& Narayan 1992, Narayan 1992).
Here we adopt an approach frequently used in non-equilibrium
thermodynamics (eg. Jou, Casa-Vasquez and Lebon 1993), and
we assume that the viscous
stress components  relax towards their  equilibrium values
on  a characteristic relaxation timescale  $\tau.$ This leads naturally to a
set of basic equations incorporating a finite
propagation speed for viscous information given
by $c\subscr{v} = \sqrt{\nu /\tau},$
where $\nu$ is the usual kinematic viscosity.

We use these to investigate the structure of the boundary layer region
between the disc and a comparatively slowly rotating star by studying
vertically averaged one dimensional models, as many of
their properties are expected to be manifested in the 
more general two dimensional case.
We begin with  a study of steady state polytropic disc models and then
go on to study
time dependent models in which energy dissipation and heat transport
are taken into account using, for illustrative purposes, parameters
appropriate to protostellar discs.

\section{Equations}
In an accretion disc the vertical
thickness $H$ is usually assumed to be small in comparison to the distance $r$
from the centre, i.e. $H/r << 1$. This is naturally expected when the
material is in a state of near Keplerian rotation. Then one can
vertically integrate the hydrodynamical equations and  work only with
vertically averaged state variables. Under the additional assumption of axial
symmetry, the vertically integrated equations of motion in cylindrical
coordinates
($r, \varphi, z$) read: 
\begin {eqnarray}
 \doverdt{\Sigma} + \frac{1}{r} \doverdr{} (r v \Sigma)
                =  0,  \label{sigma}  \\
 \doverdt{(\Sigma v)} + \frac{1}{r} \doverdr{} (r v \Sigma v)
    =
          r \Sigma \Omega^2  - \doverdr{P}
        - \Sigma \frac{G \Mstar}{r^2}
        + f_\nu
          \label{ur} \\
 \doverdt{(\Sigma r^2 \Omega)}
        + \frac{1}{r} \doverdr{} ( r \Sigma r^2 \Omega v)   =
  \frac{1}{r} \doverdr{} \left( r^2 T_{r\varphi} \right)
          \label{Om} \\
 \doverdt{(\Sigma \epsilon)}
        + \frac{1}{r} \doverdr{} ( r \Sigma \epsilon v)   =
        - \frac{P}{r} \doverdr{} ( r v ) + D_v
        - \int^{\infty}_{-\infty}\nabla \cdot {\bf F}dz
          \label{energ}
\end{eqnarray}
Here $\Sigma$ denotes the surface density 
$\Sigma = \int^\infty_{-\infty} \rho dz,$ where $\rho$ is the density.
$v$ is the radial  velocity, $\Omega$ the angular velocity, $P$ the
vertically integrated (two-dimensional) pressure,
$\Mstar$ the mass of the accreting object,
$G$ the gravitational constant, $f_\nu$ the viscous force per unit area acting
in the radial direction, and $\Trp$ is the $r\varphi$ component of the
vertically integrated viscous stress tensor. In the energy equation $\epsilon$
denotes the specific internal energy, $D_v\equiv r \Trp\doverdr{\Omega}$ is the
viscous dissipation rate per unit area , and ${\bf F}$ is the radiative energy
flux.

\subsection{Causal Viscosity}
Viscous processes are of central importance in accretion discs in that they
are responsible for the  angular momentum transport that allows the
radial inflow and accretion to occur. It is believed that processes such as MHD
turbulence are likely to be responsible for the existence of the
large viscosities, required to account for observed evolutionary timescales
associated with accretion discs (see Papaloizou and Lin 1995, and references
therein).

The essential component of the viscous stress tensor for accretion
discs is  the $( r,\varphi )$ component. The prescription normally adopted is
$\Trp = \Trpo$, where $\Trpo$ is given by an expression in the form
appropriate to a microscopic viscosity such that
\begin{equation}
    \Trpo = r \Sigma \nu \doverdr{\Omega} \label{trpo}
\end{equation}
Here $\nu$ is the kinematic viscosity coefficient. In
accretion disc theory, the $\alpha$ prescription of Shakura and Sunyaev (1973)
is often used such that 
\begin{equation}
   \nu = \alpha c\subscr{s} H,  \label{nu}
\end{equation}
Here $\alpha$ is a (usually constant) coefficient of proportionality
describing the efficiency of the turbulent transport.
In writing (\ref{trpo}) and (\ref{nu}) it is envisaged that the
turbulence behaves in such a way as to produce  a viscosity through the
action of eddies of typical size  $H$ and  turnover velocity
$\alpha c\subscr{s}.$
Vertical hydrostatic equilibrium gives
\begin{equation}
      H = \frac{c\subscr{s}}{\Omega\subscr{k}},
\end{equation}
where $\Omega\subscr{k}$ is the Keplerian angular velocity which is given by
$\Omk^2 ={GM_*/r^3}$.

The ansatz $\Trp = \Trpo$ results in the transport of angular momentum
through diffusion, with a diffusion coefficient
 $\nu \equiv \alpha \csound^2/ (\Omk)$. This
 leads formally to the possibility of instantaneous communication of
disturbances in the angular momentum distribution, or  an
infinite speed $c\subscr{v}$ of propagation of viscous information. 
In the main part of
the accretion disc this causes no serious problems, since (radial)
velocities are
very small in comparison to the sound speed. However,
in the boundary layer where
the incoming material hits the surface of the accreting object, the radial
infall velocity may become large, reaching supersonic values $|v| > \csound$
(see Papaloizou and Stanley 1986; Kley 1989; Popham and Narayan 1992).

To overcome this {\it causality} problem various rather ad-hoc
approaches that limit the viscosity in the vicinity of the star
(thus reducing the radial inflow velocity)
have been suggested (see introduction).
Here we follow a more general approach frequently used in non-equilibrium
thermodynamics (Jou et al. 1993) and also in relativistic physics
(Israel 1976) where the theory requires a finite speed of
propagation for information related to a given physical process.
One assumes that the actual turbulent
stresses tend to approach the equilibrium value $\Trpo$ 
on  a suitable relaxation time $\tau.$
This is described through an additional equation for the time evolution of
the vertically integrated $(r,\varphi)$ component of the viscous stress
\begin{equation}
  \frac{d \Trp}{dt}  = 
          {(\Trpo - \Trp)\over \tau}   \label{trp} 
\end{equation}
Note that the total or convective time derivative is used here. 

This prescription was used to model the central regions of discs
around compact objects by Papaloizou and Szuszkiewicz (1994)
who noted that the system of equations (\ref{sigma} - \ref{energ})
are then hyperbolic and thus completely causal with a propagation speed for
viscous information given by 
\begin{equation}
      c\subscr{v} = \sqrt{\frac{\nu}{\tau}}\equiv
   \csound\sqrt{\frac{\alpha}{\Omk\tau}}.
\end{equation}
Note that in the limiting case of
$\tau \rightarrow  0$, the stress is given by its equilibrium value $\Trpo.$
Also variation of $\alpha$, which may be a function of $(r,\Sigma, \Omega)$,
does not affect the causality properties of the equations.

Here we apply the above formalism using the $(r,\varphi)$ component of the
viscous stress as this is the most important for the
one dimensional models we consider. However, the formalism can be applied
to all the components of the tensor and be used in more general two
dimensional models of the type developed by Kley (1989).

\section{Steady State Polytropic models}
To illustrate, as well as simplify, we first use
a polytropic equation of state. It is found in practice that such
a treatment yields the essential behaviour of the radial and angular velocities.
We adopt
\begin{equation}
   P = K \Sigma^\gamma,
\end{equation}
where $K$ is the polytropic constant and $\gamma$ is the
adiabatic index. The local sound speed in the disc is then given by
\begin{equation} 
    \csound^2 = \doverd{P}{\Sigma} = K \gamma \Sigma^{\gamma-1}. 
\end{equation}

To analyze time independent solutions for a polytropic equation of state 
we drop the time derivatives in the evolution equations.
The continuity and angular momentum equations can then
be integrated yielding
\begin{eqnarray}
     \Mdot & = & 2 \pi \Sigma v r \\
     \Jdot & = & \Mdot r^2 \Omega - 2 \pi r^2 \Trp. 
\end{eqnarray}
Here the constants of integration denote the inward  mass flow rate $\Mdot$
through the accretion disc and the total angular momentum flow
rate $\Jdot$; both are negative.
The total angular momentum flux consists of the advective and viscous part.
Using the  radial component of the equation of motion
(assuming $f_\nu=0$) and the viscous relaxation equation,
we obtain two ordinary differential equations for $\Omega$ and $v$
(see also Papaloizou and Szuszkiewicz 1994):
\begin{eqnarray}
    \left(v^2  -  \csound^2 \right) \frac{r}{v}\frac{dv}{dr}
      & =  & \left[ r^2 \left( \Omega^2 - \Omk^2 \right) + \csound^2 \right]
             \label{dvdr}   \\    
  v\tau \left( \frac{ c\subscr{v}^2}{ v^2} -1 \right) \frac{d \Omega}{dr}  
      & =  & \left[ \Omega -
              \frac{\Jdot}{\Mdot r^2 } 
               \left( 1 - \frac{2 \tau v}{r} \right) \right]     
\label{dodr}, \end{eqnarray}
where we have also made use of the mass and angular momentum flux integrals.
We note that a complicating feature of the above 
differential equations is that they have critical points 
whenever the infall velocity reaches the sonic or viscous speed respectively.

Once values of $\alpha, \Omk\tau$ and $H/r$ have been specified, the above
system provides two first order ordinary differential equations for
$\Omega$ and $v$ with the additional parameter $\Jdot$.
Solutions can be found with $v_*$ and $\Omega$ specified at the inner boundary
with $\Jdot$ being determined as an eigenvalue in order that the exterior
solution matches onto a Keplerian disc. We present here results
for illustrative examples with  $\alpha=0.01,$ 
and $\Omk\tau$ in the range $0.1-25.$ In all cases $\Omega/\Omk$
was taken to be one third at the inner boundary, $\gamma =2,$ and
$H/r \sim 0.05$ in the Keplerian part of the disc.
Each model has a constant value of $c\subscr{v}/\csound.$
Details of the models are given in table 1.
\begin{table}
\caption[]{Parameter of the stationary polytropic and time dependent
radiative models}
\begin{tabular}{lllllclllll}
\multicolumn{5}{c}{Polytropic Models} & \hspace{1.5cm} & 
  \multicolumn{5}{c}{Radiative Models}\\
\cline{1-5} \cline{7-11}
Nr. & $\alpha$ & $\Omega_k\tau$ & $c\subscr{v}/\csound$ & $v_*$ 
& \hspace{1.5cm} & Nr. & $\alpha$ &
    $\Omega_k\tau$ & $c\subscr{v}/\csound$ & Remarks  \\
\cline{1-5} \cline{7-11}
1 & 0.01 & 0.1 & 0.316 & 0.1   &  &  
  11 &  0.01 & 1 & 0.10 & stable, with overstab. \\
2 & 0.01& 1.0 & 0.1 & 0.01     &  & 
  12 & 0.01 & 25 & 0.02 & stable, overstab. damped \\ 
3 & 0.01 & 4.0 & 0.05 & 0.01   &  & 
   13  & 0.1 & 10 & 0.10 & stable, $\Mdot=10^{-6}$\\
4 & 0.01 & 9.0 & 0.033 & 0.01  &  & 
   14  &  0.1 & 250 & 0.02 & stable \\
5 & 0.01 & 25.0 & 0.02 & 0.01  &  &  
   15  &  0.1 & 10 & 0.10 & unstable \\
  &      &      &      &       &  &  
   16  &  0.1 & 0.4 & 0.50 & unstable  \\
\end{tabular}
\label{results}
\end{table}
All of our calculations are such that the flow remains subviscous throughout.
In all cases, except perhaps model 1, which has the largest value
of $c\subscr{v}/\csound$ for $\alpha =0.01,$ the flow  speed almost  reaches
the viscous speed at its maximum. In such cases the $\Omega$ profile 
becomes nearly discontinuous (Fig. 1a).  For model 1, the profile
is moderately extended approximately up
to the pressure scale height in the slowly rotating star.
However, in models 2-5 the profile approaches a discontinuity.
The jump in angular velocity occurs when the flow speed is at a maximum
and almost equal to the viscous speed (Fig. 1b).
At the discontinuity, there
is a jump in the velocity gradient.
The tendency to form a discontinuity is even more noticeable 
in models which have $\alpha=0.1$ (not shown).
\begin{figure*}[htb]
      \vbox{\psfig{figure=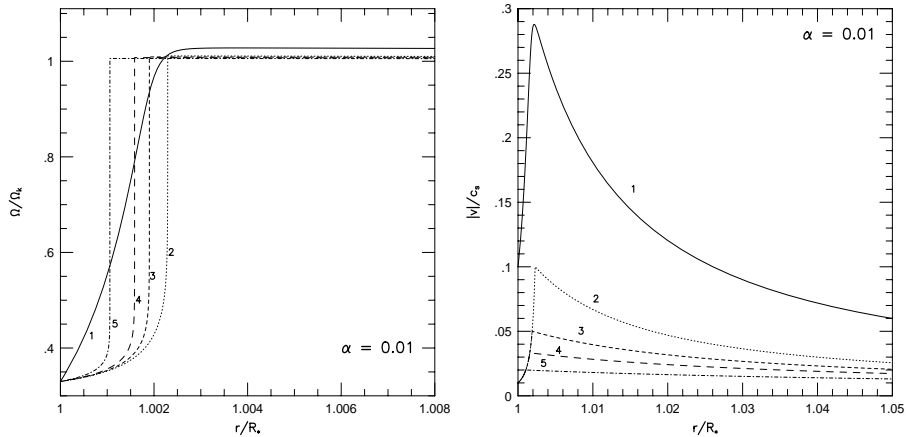,width=12.2cm,%
          angle=-90.,clip=}}\par
      \caption[]{Ratio of the angular velocity $\Omega$ to the Keplerian
value $\Omega_k$ and radial Mach number $|v|/c_s$ versus radius for
models 1 to 5.}
      \label{fig1}
\glossary{object 3} 
\end{figure*}
 We note that the radial
equation of motion (\ref{dvdr})
implies that a discontinuity in $\Omega$ must occur
at constant velocity and be accompanied by a jump in velocity gradient.
The occurrence of these near discontinuities is reminiscent of the `shear
shocks'
\glossary{shear shock}
envisaged by Syer and Narayan (1993). At such locations material
is instantaneously slowed down as it encounters the stellar surface.
 However, they occur
here at the viscous speed only and do not involve a transition  from
super to subviscous speeds. The discontinuities tend to be approached
whenever the model is strongly affected by the causality condition.
The condition $R_*/H  >  c\subscr{v}/(\alpha \csound)$ provides a rough
indication as to when this occurs in the models presented in this section

\section {Time dependent calculations}
In addition to the steady state calculations described  above, we also
studied time dependent evolution of  the flow in order to investigate any 
essential unsteady behaviour associated with the boundary layer.
The numerical solution of equations (\ref{sigma}) to (\ref{energ}) is
accomplished by a finite difference method. 
The partial differential equations are discretized on a spatially fixed
one-dimensional grid that stretches from $r=1$ to $r=2$.
This computational domain is covered by typically 1000 grid cells.
A forward time and centered space method with operator
splitting and a monotonic advection scheme is used (Kley 1989).

We have studied polytropic models of the type used above in which the energy
equation (\ref{energ}) is dropped, as
well as models which include (\ref{energ}) with heat transport.
For these cases
\begin{equation} 
  \int^{\infty}_{-\infty}\nabla
\cdot {\bf F}dz =  2\sigma T\subscr{eff}^4 
                 - {H/r}{\partial\over \partial r}( rF_r),
\end{equation}
where $T\subscr{eff}$ is the effective temperature at the disc surface,
$\sigma$ is Stefan's constant, and $F_r$ is the radial radiative flux.
For our models we used an analytic approximation to tabulated opacities
(Lin \& Papaloizou 1985).
The gas consists of a Hydrogen and Helium mixture where
the dissociation of $H_2$ and the degrees
of ionization of H and He are calculated by solving the Saha equation. 

We adopted conditions appropriate to protostellar discs, where the protostar
has $M_* = 1.0\Msol, R_* = 3R_{\odot}$, and $\Tstar=4000K$.
Through the surrounding disc, a mass flow rate of
$\Mdot =10^{-7} M_{\odot} yr^{-1}$ is accreted (only model 13 has
$\Mdot =10^{-6} M_{\odot} yr^{-1}$).
At the inner boundary a fixed outwardly
directed stellar flux, $F_* = \sigma T\subscr{*}^4$, is assumed.
The radial infall velocity at $R\subscr{min}=R_*$
is fixed at a given small fraction of local Keplerian velocity at
$R_*$. We use typically $v_* = 10^{-3} v\subscr{k*}$. The stellar
angular velocity is $0.3$ of the break-up velocity
for the polytropic test cases,
and to $0.1$ for the fully radiative models. 
 
At the outer boundary the angular velocity is Keplerian, the radial radiative
flux vanishes and the radial infall velocity and the density are prescribed
in such a way to ensure a given constant mass inflow rate through the system.
For initial conditions we use a simple polytropic disc model
with no boundary layer.
The system is subsequently evolved until 
the region containing the boundary layer attains  a quasi-steady state 
which was  reached in most cases.
Oscillations caused by viscous overstability
persist typically near the outer boundary (see Kato 1978, Godon 1995).

In order to compare with the steady state calculations described 
above we have considered time dependent polytropic models with
constant $\alpha,$ and $\Omk \tau.$
There was, in general, positive agreement between the two methods.
There is a tendency
for the evolutionary calculations to overshoot the viscous speed somewhat,
an effect which decreases with increasing spatial resolution
of the calculations.

\subsection{Radiative models}
The parameters of the calculations with thermal effects included
are listed in Table~1.
Solutions appropriate to a statistically steady state are 
presented for models (14, 15) which both have $\Mdot =10^{-7} M_{\odot},$
and $\alpha =0.1$.
The viscous velocity $c\subscr{v}=\csound\sqrt{\alpha/(\Omk\tau)}$
differs by a factor of $5$ between the two models.
Some state variables are plotted in figure $2$.
\begin{figure*}[htb]
      \vbox{\psfig{figure=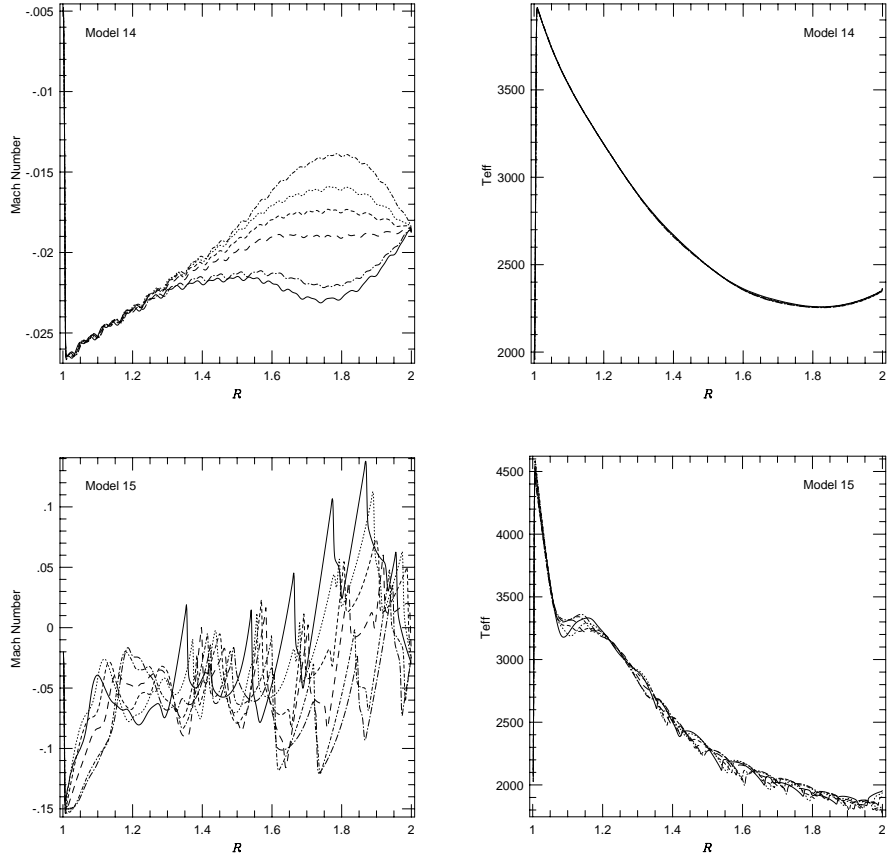,width=12.2cm,%
          angle=00.,clip=}}\par
      \caption[]{The radial mach number and the effective temperature
versus radius for models 14 and 15 (see table 1) at six different
times.}
      \label{fig2}
\glossary{object 3} 
\end{figure*}
The structure of the $v$ and $\Omega$ profiles is similar to that in the
polytropic case, i.e. $\Omega$ displays a near discontinuity and
$v$ has a peaked maximum near the viscous speed.
The rate of liberation of energy in the boundary layer  is
$0.5 \Mdot R_*^2\Delta\Omega^2,$ with $\Delta\Omega$ being the jump in
$\Omega$ that occurs there. 
Heat diffusion then  occurs over a greater
length scale.
In the case with $\Omega\tau = 25,$ the optical depth is about ten times
larger than that with  $\Omega\tau = 1.$  But this only results  in a $20$
percent reduction in $T\subscr{eff}$  because the latter
quantity is predominantly
determined by a fixed  rate of energy production.
However, typically  there will be an order of magnitude difference in
the estimated value  of $\Mdot$ for which the boundary layer becomes optically
thin. This is
the main effect
of increasing the relaxation time $\tau.$ Another consequence of
increasing $\tau$ is  damping of the present viscous overstability because
of the stronger phase lag induced. In model 12 the overstability is
eventually damped completely.
We ran three models with $\alpha=0.1$ which had constant values
of $\Omk\tau$ chosen such that $c\subscr{v}/\csound$ was $0.02, 0.1$ and $0.5$,
respectively.
The case with $c\subscr{v}/\csound=0.02$ (Model 14, figure 2) behaved
in a very similar way to the
cases with $\alpha =0.01$, in that it had an almost steady and 
stable inner boundary layer region.  
\begin{figure*}[htb]
      \vbox{\psfig{figure=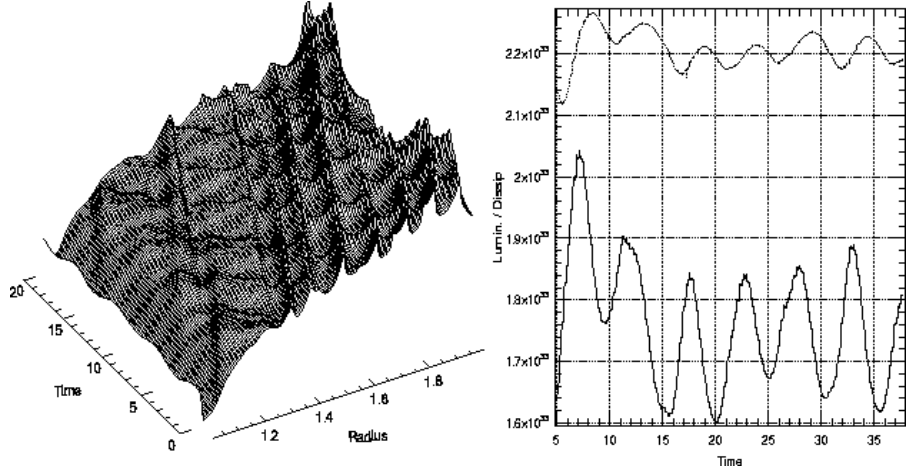,width=12.2cm,%
          angle=-90.,clip=}}\par
      \caption[]{Time variation of the density for model 15 (a).
Luminosity (dashed line) and total dissipation (solid line) versus time (b).
value $\Omega_k$ and radial Mach number $|v|/c_s$ versus radius for
the same model. Time given in dimensionless units with an offset.}
      \label{fig3}
\glossary{object 3} 
\end{figure*}
However, highly unstable features remained in the boundary layers
of the other two models (15, 16) with $\alpha =0.1.$ These features
were not seen
in the other models even when they showed signs of viscous overstability.
 Here the characteristic wavelength is much shorter and the temperature
structure and power output are significantly affected in a highly
irregular manner.
The instability appears more pronounced with higher $c\subscr{v}/\csound$, here
the strongest instabilities are present for $c\subscr{v}/\csound=0.5$.
Then slightly supersonic speeds with shocks, as well as
superviscous speeds occur.
We remark that in the models with
$\alpha =0.1$, the short wavelength inward coming compressional waves are not
expected to be reflected at a Lindblad resonance
  before they reach the boundary layer region, so they are able to
affect the power output there.  This would indicate that if the outer boundary
condition allows,  waves may exist 
in the boundary layer region where they may significantly
affect the power output. In Figure 3b the variation of the luminosity
and the total dissipation is shown over a short time interval for model 15.
The luminosity varies over timescales of the order of the dynamical
time (orbital Keplerian period at the stellar surface) but the variations
in amplitude are less than 4\%.

Note  that the driving mechanism for these motions is not just simple
viscous overstability which was never seen in the boundary layer
region. The generation of superviscous and sometimes supersonic speeds
indicates that the nature of the causal description must play a role.
In figure 3a the time variation of the density is displayed in a
three-dimensional space time diagram. It is clear that there are
quasi-periodic 
wave-like perturbances moving from the inside outwards which are generated
in the vicinity of the boundary layer. The perturbations
interact intricately with reflected waves moving inwards (Fig. 3a).

These waves occur in a region where the central temperatures
lie somewhat below $10^4K$, where opacity rises rapidly with temperature.
Hence, even though the central temperatures vary only very little, the
effective temperature displays much stronger variations. The origin
of the disturbances lies in an interaction of the radiative transport
with the causal viscous transport. For higher $|v/c\subscr{s}|$, the
interaction
becomes very much stronger leading to variations in luminosity of 
a few percent in the case of model 15.
Notice that models 13 and 15 have identical parameter
$\alpha, \tau$, and $c\subscr{v}/c\subscr{s}$, and only differ in the mass
inflow rate which is ten times higher in model 13. The increased disc
thickness for the higher $\Mdot$ model leads to a larger optical depth
and higher temperatures that drive the system out of the instability region.

%
\vspace{.2cm}\noindent{\it Acknowledgement:}
W.~K. would like thank the Astronomy unit  at QMW for their kind
hospitality during two visits. This work was supported by the EC
grant ERB-CHRX-CT93-0329.

\end{document}